\documentclass[10pt,journal,compsoc]{IEEEtran}
\ifCLASSOPTIONcompsoc
  \usepackage[nocompress]{cite}
\else
  \usepackage{cite}
\fi

\usepackage{amsmath,amssymb,amsfonts}
\usepackage{multirow}
\usepackage{graphicx}
\usepackage{textcomp}
\usepackage{xcolor}
\usepackage{color}

\usepackage{array,multirow,makecell}
\usepackage{mathrsfs}
\usepackage{float}
\usepackage{graphicx}
\usepackage{amsmath}
\usepackage{array}
\usepackage[english]{babel}
\usepackage{mdwmath}
\usepackage{mdwtab}
\usepackage{tabularx}
\newcolumntype{P}[1]{>{\centering\arraybackslash}p{#1}}
\newcolumntype{M}[1]{>{\centering\arraybackslash}m{#1}}
\usepackage{amssymb}
\usepackage{textcomp}
\usepackage{booktabs}
\usepackage[justification=centering]{caption}
\usepackage{amsthm}
\theoremstyle{definition}
\newtheorem{definition}{Definition}[section]
\newtheorem{theorem}{Theorem}[section]

\newcommand{\etal}{\hbox{{et al.}}\xspace}
\newcommand{\eg}{\hbox{{e.g.}}\xspace}
\newcommand{\ie}{\hbox{{i.e.}}\xspace}

\newcommand{\etc}{\hbox{{etc.}}\xspace}

\usepackage[linesnumbered, ruled]{algorithm2e} 
\usepackage{algpseudocode}

\SetAlFnt{\small}
\SetAlCapFnt{\small}
\SetAlCapNameFnt{\small}
\SetAlCapHSkip{0pt}
\IncMargin{-\parindent}
\newcommand{\Fname}{\textsc{FedForest}\xspace}

\SetCommentSty{mycommfont}

\newenvironment{packeditemize}{
\begin{list}{$\bullet$}{
\setlength{\labelwidth}{8pt}
\setlength{\itemsep}{0pt}
\setlength{\leftmargin}{\labelwidth}
\addtolength{\leftmargin}{\labelsep}
\setlength{\parindent}{0pt}
\setlength{\listparindent}{\parindent}
\setlength{\parsep}{0pt}
\setlength{\topsep}{3pt}}}{\end{list}}

\begin{document}
\title{
An Interpretable Federated Learning-based Network Intrusion Detection Framework
}

\author{Tian~Dong,
        Song~Li, 
        Han~Qiu,
        and~Jialiang~Lu
        
\IEEEcompsocitemizethanks{
\IEEEcompsocthanksitem T. Dong and S. Li, and J. Lu are with Shanghai Jiao Tong University, Shanghai, China, 200240. Email: tian.dong@sjtu.edu.cn, antoinels.cn@gmail.com, jialiang.lu@sjtu.edu.cn.
\IEEEcompsocthanksitem H. Qiu is with Institute for Network Sciences and Cyberspace, Beijing National Research Center for Information Science and Technology, Tsinghua University, Beijing, China, 100084. Email: qiuhan@tsinghua.edu.cn.
}
\thanks{Manuscript received xxx.}
}

\markboth{}%
{Shell \MakeLowercase{\textit{et al.}}: Bare Demo of IEEEtran.cls for Computer Society Journals}


\IEEEtitleabstractindextext{%
\begin{abstract}

Learning-based Network Intrusion Detection Systems (NIDSs) are widely deployed for defending various cyberattacks. 
Existing learning-based NIDS mainly uses Neural Network (NN) as a classifier that relies on the quality and quantity of cyberattack data. 
Such NN-based approaches are also hard to interpret for improving efficiency and scalability. 
In this paper, we design a new local-global computation paradigm, \Fname, a novel learning-based NIDS by combining the interpretable Gradient Boosting Decision Tree (GBDT) and Federated Learning (FL) framework. 
Specifically, \Fname is composed of multiple clients that extract local cyberattack data features for the server to train models and detect intrusions. 
A privacy-enhanced technology is also proposed in \Fname to further defeat the privacy of the FL systems. 
Extensive experiments on 4 cyberattack datasets of different tasks demonstrate that \Fname is effective, efficient, interpretable, and extendable. 
\Fname ranks first in the collaborative learning and cybersecurity competition 2021 for Chinese college students.

\end{abstract}

}

\maketitle

\IEEEdisplaynontitleabstractindextext

\IEEEpeerreviewmaketitle

\section{Introduction}
\label{sec:introduction}
As one of the most important infrastructures of the Internet, Network Intrusion Detection Systems (NIDSs) are designed to detect malicious behaviors or intrusions in the Internet~\cite{AttackOnKitsune}. 
As a gatekeeper for Internet devices, NIDS plays an important role in defeating cyber-attacks for enterprises, individuals, and governments on various computing systems~\cite{humayun2020cyber}.
Nowadays, deploying the Machine Learning (ML) methods to improve the performance of NIDSs become a promising direction. 
Benefiting from the extraordinary ability to extract and learn features from existing network data, these learning-based NIDSs are proved to outperform many existing NIDSs such as~\cite{DBLP:journals/jaihc/SafaldinOA21,DBLP:journals/fi/FerragMADJ20,DBLP:journals/js/LiYBCC18,DBLP:conf/ndss/MirskyDES18}.

The current popular trend of building a learning-based NIDS is to train an anomaly detection model based on benign network traffic. 
This anomaly detection model can characterize the normal behaviors of network traffic and identify intrusions by calculating the deviations from the normal behaviors. 
The output of such models, such as Kitsune~\cite{DBLP:conf/ndss/MirskyDES18}, is an anomaly score that will generate an intrusion alarm when this score exceeds a pre-set threshold. 
Without requiring any known attack signatures, such systems can effectively generate alarms for various kinds of intrusions including even unknown attacks which have distinct behaviors from the normal traffic. 

However, such an anomaly detection-based NIDS approach still has several obvious limitations. 
(1) The accuracy of these learning-based NIDS highly relies on the quality and quantity of the training data which is not suitable for the data privacy-sensitive scenarios. 
(2) The output of such NIDS is just an anomaly alert but still requires human experts to analyze and identify to understand the attack category to react. 
(3) Most of such NIDSs are designed based on ML models that are non-interpretable for further improvement and robustness. 
(4) These NIDSs lack scalability since when dealing with different anomaly detection tasks, significant modifications on the ML models are usually required for training a new model. 

In this paper, we propose \Fname, \textit{a novel privacy-preserving, interpretable, and scalable NIDS}, to solve the above limitations. 
First, \Fname uses a Federated Learning (FL) framework to solve the data privacy issue. 
A client-to-server architecture is deployed in our work to let the client share the privacy-preserving information of various kinds of network traffic data for the NIDS training on the server. 
Enhanced privacy protection is given to further protect the data against the attacks like DLG~\cite{deep_leakage}.
Second, \Fname can classify the exact attack category for immediate reaction. 
Also, our framework can support the extension for adding the unknown attacks by letting the client upload the novel attack data update the classifier with little cost. 
Third, \Fname chooses Gradient Boosting Decision Tree (GBDT) model as the core classification algorithm. 
GBDT, as a tree-based ML method, is interpretable and efficient compared with the other ML-based approaches like NNs~\cite{DBLP:conf/icissp/SharafaldinLG18a, sustainable_ensemble_nids}. 
In the end, for scalability, \Fname can achieve high accuracy on different intrusion detection tasks without modifying the system architecture and model structures. 

We conduct extensive experiments for \Fname on $4$ cyberattack datasets for diverse cybersecurity tasks, including Distributed Denial-of-Service (DDoS) attack classification task (DDoS2019~\cite{cicddos19}), DNS over HTTPS (DoH) traffic classification task (DoHBrw2020~\cite{DoH}), darknet classification task (Darknet2020~\cite{darknet}), and Android malware classification task (MalDroid2020~\cite{maldroid}). 
The results indicate that \Fname can achieve a high accuracy supported by the FL framework with data privacy protected. 
We show that \Fname is interpretable which can be easily used for human experts' intervene and for further improvement~\cite{meng2020interpreting}. 
Also, \Fname is able to alert for unknown attacks. 
In addition, \Fname can be efficiently trained and is scalable for providing high accuracy on these different intrusion detection tasks. 

In summary, our contributions can be summarized as follows. 
(1) We propose a novel collaborative learning framework, \Fname, by combining FL and GBDT to build high accuracy NIDS. 
(2) \Fname can solve the limitations of existing learning-based NIDS especially the anomaly detection methods by giving the exact attack category with interpretability and scalability. 
(3) With extensive evaluation, \Fname is proved to be effective considering scalability, efficiency, and interpretability on three well-known network intrusion datasets and one Android malware dataset. 

This paper is organized as follows. 
In Section~\ref{sec:background}, we compare previous related works on NIDSs and introduce basic notions of FL and ML models like GBDT and NNs. 
In Section~\ref{sec:problem}, we present an overview and several challenges for FL application on NIDSs, and the problem settings of our study. 
In Section~\ref{sec:approach}, we show \Fname training and inference algorithms in detail. 
In Section~\ref{sec:experiments}, we demonstrate the superiority of \Fname by various experiments on different cyberattck classification datasets. We discuss and list future work in Section~\ref{sec:discuss} and conclude in Section~\ref{sec:conclusion}.

\section{Background}
\label{sec:background}
In this section, we firstly present FL, including the definition and principles of horizontal and vertical FL. 
Then, since most NIDSs adopt tree-based ML models and NNs as the backbone classifier for NIDSs, we provide an overview of previous works of NNs-based and tree-based NIDSs, analyze their advantages and limitations, and present our research motivation. 
We conclude this section by introducing differential privacy~\cite{dwork2006calibrating} which is widely used for enhancing privacy protection for the FL framework.

\subsection{Federated Learning}

Federated Learning (FL)\cite{DBLP:journals/corr/KonecnyMYRSB16} is a distributed learning paradigm that allows training model with data of other participants without data sharing. 
FL enables a model to capture more features and enhance generalization while keeping data privacy. 
In practice, the FL setting comprises a server that holds a model and several clients holding the private data. 
The server coordinates clients to train the model while preserving client data locality, i.e. no transfer required for the client's private data. 
Specifically, the server distributes the model to clients to train the model with their own data and collect back the updated weight. 
According to the data feature distribution on clients, FL can be categorized into Horizontal FL and Vertical FL. 
FL can be deployed in the scenario that data contains private information that is hard to collect and share such as the cyberattack data. 
Combining NIDS with FL can let the NIDS classifier learn more cyberattack features from many parties which can help build effective learning-based NIDS.

\begin{figure}[t]
    \centering
    \includegraphics[width=0.95\linewidth]{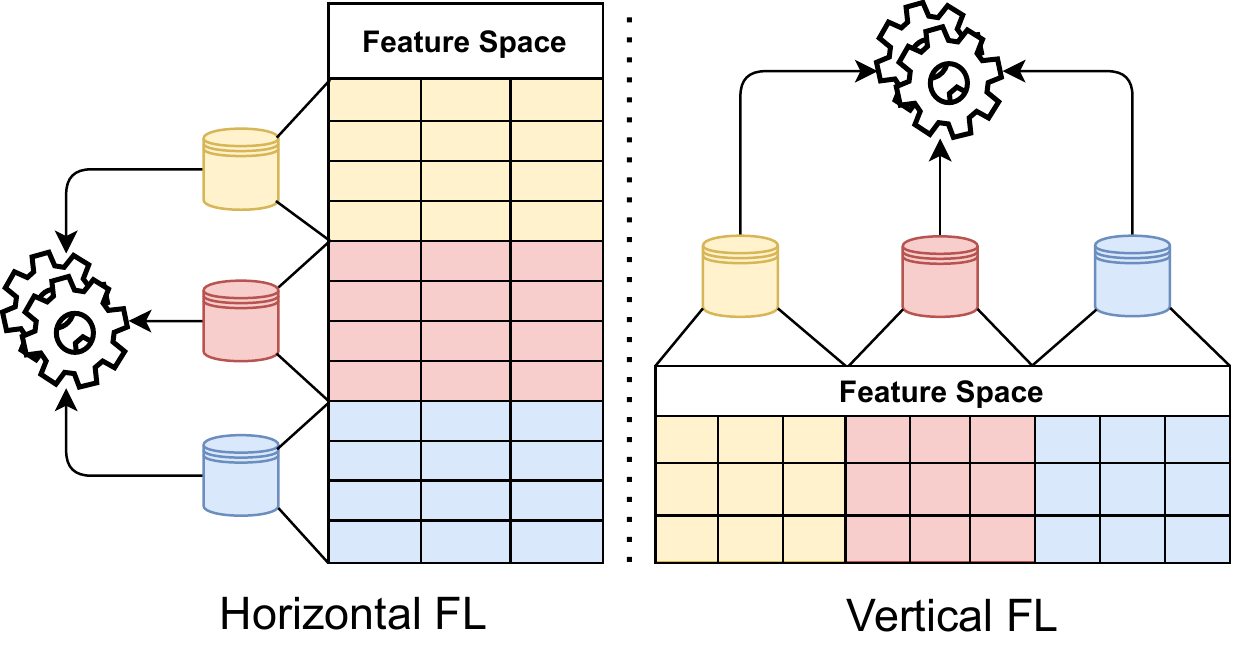}
    \caption{Illustration of HFL and VFL.}
    \label{fig:flconcepts}
\end{figure}


Vertical FL, or feature-wise FL, whose architecture is shown in \figurename~\ref{fig:flconcepts}, is used for scenarios where clients share the same sample space. For example, Vertical FL can be applied for FL between banks and e-commerce companies: banks and e-commerce companies share the same client group but have different features of clients.

Horizontal FL, or sample-wise FL, is used for scenarios where clients' data share the same feature space. 
The architecture is shown in \figurename~\ref{fig:flconcepts}. 
An illustrative example can be an object detection FL training task: server trains global model with image data located on users' smartphone. 
Existing FL frameworks are mainly designed for training NNs model under horizontal FL setting, such as FedAvg \cite{federated_learning}.
In FedAvg, instead of running only one SGD step, each client takes several SGD steps locally, and then sends their parameters to the server for update aggregation. 
Other frameworks~\cite{Fedprox, SCAFFOLD, MOON} optimized local updating and gradient aggregation, resulting in higher accuracy score. 
Horizontal FL can also be applied for DDoS attack detection~\cite{DBLP:conf/icdcs/NguyenMMFAS19}.
The server uses the statistic data of network traffic gathered by the routers, \ie the FL clients, to train the NIDS classifier. In general, cyberattack data share the same feature space and differ on the sample space. 
We only consider the horizontal FL setting in this paper.

\noindent{\textbf{FL-based NIDSs.}}
There are several works~\cite{DBLP:conf/icdcs/NguyenMMFAS19, deepfed2020, DBLP:journals/iotj/LiuGNZXKH21} that proposed FL-based NIDSs. 
Nguyen~\etal~\cite{DBLP:conf/icdcs/NguyenMMFAS19} proposed D\"iot, an autonomous FL-based anomaly detection system for IoT devices. 
Liu~\etal~\cite{DBLP:journals/iotj/LiuGNZXKH21} designed an anomaly detection framework with FL and DNN for time-series data in IoT. 
Cetin~\etal~\cite{deepfed2020} proposed an FL framework for NIDSs in industrial networks.
However, they only focus on anomaly detection tasks and their backbone models are DNNs-based without considering the interpretability. 

\subsection{NN-based NIDS}

DNNs are widely used for tasks of CV and NLP because of their extraordinary ability of high-dimensional feature extraction. 
In this paper, we only consider Multi-Layer Perceptron (MLP) as cybersecurity data, \eg network traffic data which are mostly tabular data of low dimension (data of less than $100$ features). 
Concretely, a $n$-layer MLP $F$ is composed of a set of weight matrix $\{W_i\}_{1\leq i\leq n-1}$ and a bias vectors $\{b_i\}_{1\leq i\leq n-1}$. 
For each $1 \leq i\leq n-1$, we have $W_i\in \mathbb{R}^{n_{i+1}\times {n_i}}$ and $b_i\in\mathbb{R}^{n_{i+1}}$. 
With input $\mathbf{x}\in \mathbb{R}^{n_1}$, $F(\mathbf{x})=F_{n-1}\circ F_{n-2}\circ \dots \circ F_1(\mathbf{x})$, where $F_i(\mathbf{x}) = f(W_i\mathbf{x}+b_i)$, $f$ is the activation function and we choose ReLU for the baseline model in this paper. 
Note that before feeding data into NNs, it is necessary to preprocess data in order to stabilize training and optimize performance. 
We will clarify preprocessing in Section~\ref{sec:experiments}. 

To apply NNs in NIDSs, researchers seek to come up with more suitable NNs architectures. 
For instance, the state-of-the-art NIDS, Kitsune~\cite{DBLP:conf/ndss/MirskyDES18}, is a plug-and-play NIDSs for IoT devices based on NNs. 
The key part of Kitsune, the classifier, is a $2$-layer autoencoder variant.
Sasanka~\etal~\cite{potluri2018convolutional} apply Convolutional Neural Networks (CNNs) in their multi-class NIDSs and evaluate DDoS detection tasks.
Riyaz~\etal~\cite{DBLP:journals/soco/RiyazG20} build a new CNNs-based intrusion detection model and demonstrate its effectiveness on wireless network data. 
Wen~\etal~\cite{DBLP:journals/ijahuc/WenSDKR21} introduce deep belief networks into their intrusion detection model for wireless sensor networks. 
Xu~\etal~\cite{meta_learning_nids} leverage meta-learning to devise a few-shot learning NIDS.

\subsection{Tree-based NIDSs}

\begin{figure}[t]
    \centering
    \includegraphics[width=0.7\linewidth]{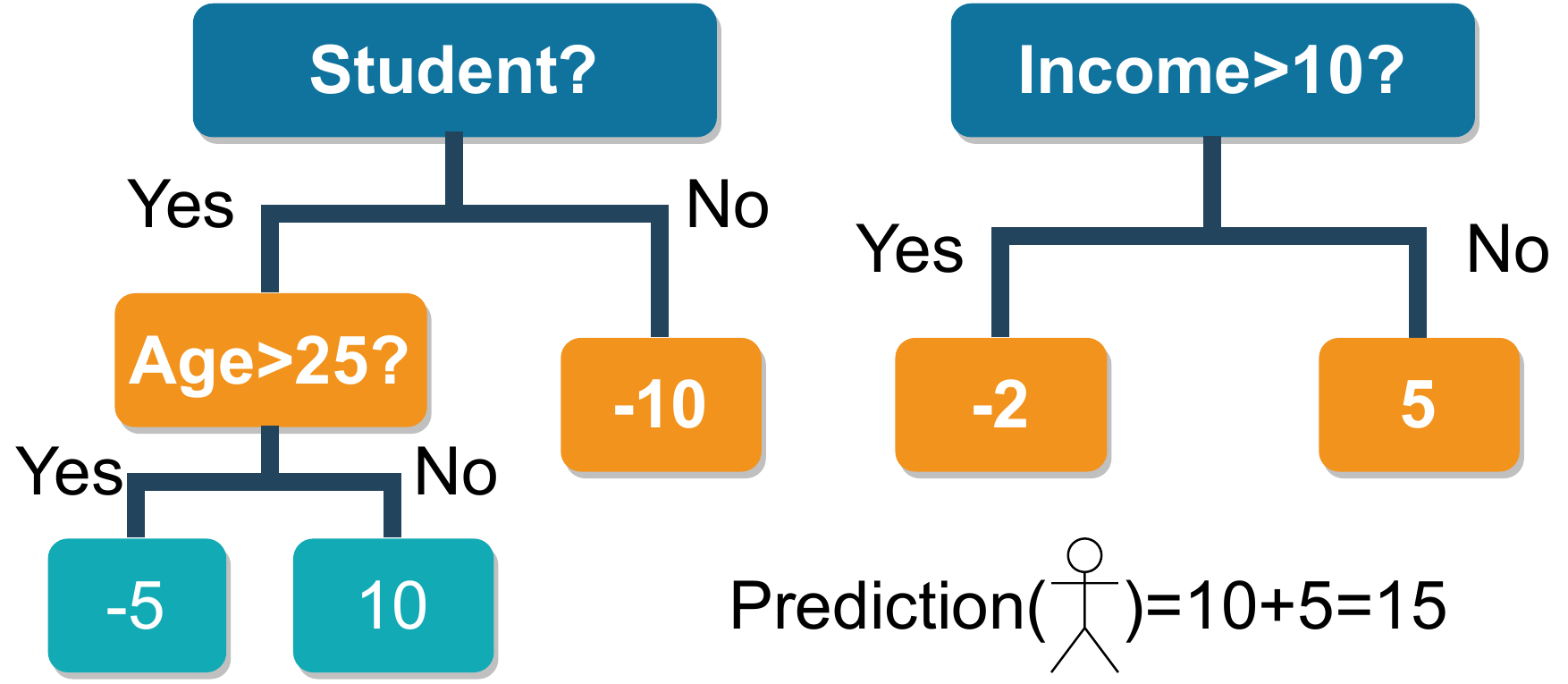}
    \caption{An example of GBDT model.}
    \label{fig:gbdt_example}
\end{figure}

A decision tree is a tree-based interpretable ML model composed of a root node and a set of inner nodes and a set of leaf nodes. 
Each node represents a learned decision condition, \eg the $i$-th feature of input is larger than $0$, and nodes are connected according to the learned decision rules. 
The inputs are firstly fed to the root node, and next to an inner node if the decision condition of the root node is satisfied, or another inner node otherwise. 
GBDT is an ensemble of decision trees that are learned by minimizing learning loss function which is determined by tasks like classification or regression. 
Specifically, for a GBDT model with $T$ decision trees, the prediction score or the logit of input $\mathbf{x}$ is the sum of output of decision trees:

\begin{equation}
    \hat{y} = \sum_{t=1}^T f_t(\mathbf{x}),
\end{equation}

where $f_t$ is the $t$-th decision tree. 
An example is shown in \figurename~\ref{fig:gbdt_example} where a student data sample is scored $10$ and $5$ by $2$ decision trees, respectively. 
The output logit of GBDT is the sum $15$. 
For the classification task, GBDT model first computes a logit score for each class by an equal number of class-dependant decision trees. Then, the probability of each class is computed which is the softmax of the logit scores. 

There are numerous works that incorporate decision trees to NIDSs. 
Tulasi~\etal~\cite{RF_DT_NIDS2020} mix random forest and decision trees to build an anomaly-based NIDS.
Li~\etal~\cite{DBLP:journals/js/LiYBCC18} use Particle Swarm Optimization (PSO) algorithm to optimize the proposed hybrid model and obtain better results over their compared methods.
The RDTIDS\cite{DBLP:journals/fi/FerragMADJ20} combines decision tree and rules-based concepts and achieves state-of-the-art binary classification results.
The NIDS proposed by Ranjit~\etal~\cite{panigrahi2021consolidated} adopts Consolidated Tree Construction (CTC) algorithm with a C4.5-based detector, obtaining a nearly perfect accuracy score on DDoS attack datasets.

Even though numerous models have been proposed, we observe that only few of them~\cite{potluri2018convolutional, sustainable_ensemble_nids} are designed for multi-class attack detection. 
However, both of them are not considered to be scalable for other cyberattack detection tasks.
Besides, model interpretability is also decisive for security-critical applications. 
On the one hand, model interpretability could help to improve the model robustness to mitigate issues like Adversarial Examples (AE)~\cite{DBLP:journals/corr/GoodfellowSS14}. 
On the other hand, interpreting a model could improve its performance as shown in~\cite{meng2020interpreting}. 
Thus, deploying a tree-based approach to build NIDS has obvious advantages over those NN-based NIDSs which are hard to include interpretability. 



\subsection{Privacy Enhanced Technology}
FL has been regarded as an optimal solution for distributed model training without private data sharing. 
However, it has been demonstrated that by using the difference between model parameter updates, insider attackers can recover sensitive information or even original training data of a small training batch~\cite{deep_leakage,iDLG}. 
Recently, Yin~\etal~\cite{yin2021see} propose a strong privacy attack that can recover training data of a large batch. 
Therefore, using privacy-enhancing techniques for FL scenario is necessary. 

Data augmentation techniques are proved to be effective for data privacy in FL training~\cite{gao2020privacy} but are only limited to computer vision FL tasks. 

To protect NNs models, data augmentation techniques, \eg image mixup, are effective for defending gradient-based privacy attack~\cite{gao2020privacy}, but such techniques are only limited to computer vision tasks. 
More privacy-enhancing techniques include differential privacy~\cite{dwork2006calibrating} and homomorphic encryption~\cite{gentry2009fully}, and have been applied in general FL frameworks to provide data privacy~\cite{privateFDL2020TDSC,personalizeFLprivacy,wei2020federated}. 
The FL framework NbAFL~\cite{wei2020federated} proposed by Wei~\etal add artificial noise on the clients before model parameter aggregation, and the convergence of training has been theoretically proved. 
Hu~\etal~\cite{personalizeFLprivacy} combine personalized model and differential privacy in their FL framework of personalized services. Xu~\etal~\cite{privateFDL2020TDSC} use Yao's garbled circuits and homomorphic encryption systems to provide the data confidentiality of their FL framework. 

Data masking is a privacy-enhancing method that protects raw data by randomly dropping a proportion of data before training starts. 
For example, Qiu~\etal~\cite{masking_lr} use data masking to ensure the privacy and efficiency of their linear regression model. 
Differential privacy provides guaranteed privacy protection and can be also be efficiently implemented. 
In the next, we introduce the basic definition of differential privacy, and Laplace Mechanism~\cite{dwork2006calibrating}, a method to achieve differential privacy of prefixed privacy budget.    

Differential privacy provides provable privacy guarantees for data exchange under FL settings. The main idea is to make the model's outputs similar given two adjacent datasets. We show the formal definition as below.

\begin{definition}[$\epsilon$-differential privacy]
A randomized function $F$ guarantees $\epsilon$-differential privacy if 
\begin{equation}
\mathbb{P}(F(D)\in S) \leq \exp(\epsilon) \mathbb{P}(F(D')\in S)
\end{equation},
where $D$ and $D'$ are two datasets that differ on only one sample, $\epsilon$ is privacy budget.
\end{definition}

The privacy budget measures the privacy-utility trade-off: A smaller $\epsilon$ provides a stronger privacy guarantee.
Denote the sensitivity of $F$ as $\Delta(F)  := \max_{(D,D')\in\mathcal{D}^2}\Vert F(D)-F(D') \Vert_1$, where $\mathcal{D}$ is the set of all possible datasets. We use Laplace Mechanism~\cite{dwork2006calibrating} to achieve $\epsilon$-differential privacy as the data to be protected are numerical.

\begin{theorem}[Laplace Mechanism]
Let $F:D\rightarrow \mathbb{R}^d$ be a numerical function, the Laplace Mechanism $\mathcal{M}$ for $D\in\mathcal{D}$ 
\begin{equation}
\label{eq:lap_mec}
    \mathcal{M}(D) = F(D) + X_{Lap},
\end{equation}
where $X_{Lap}$ is noise drawn from Laplace distribution with mean $0$ and scale $\frac{\Delta F}{\epsilon}$, provides $\epsilon$-differential privacy.
\end{theorem}

\section{Problem Statement}
\label{sec:problem}
In this section, we present the problem statement to illustrate our research motivation and our research goals. 


The problem that we aim to solve can be described with the following description. 
One cybersecurity service provider that plans to set up a NIDS for multi-class cyberattack detection as a service for his clients. 
In order to achieve such a goal, the cybersecurity service provider must build a model based on the network traffic data and features collected by his clients. 
However, the clients, such as small enterprises or governments, cannot directly provide this necessary network traffic data since the data privacy cannot be sacrificed. 
Then, the NIDS must be able to automatically identify the cyberattack category since we assume the clients such as small enterprises or governments are lack security experts that can immediately react based on only an anomaly alarm. 
Third, the NIDS must be interpretable due to the reason that interpretability can help to improve the NIDS performance in the future and potentially provide robustness. 
In the end, once the NIDS is built, the cybersecurity service provider can also extend the usage of NIDS to diverse cyberattack detection tasks without too much modification on the NIDS including the system architecture and ML model structures. 


In this paper, we aim to develop a novel learning-based NIDS for multi-class intrusion classification. 
This NIDS should provide high accuracy while keeping the data locality and privacy. 
Also, NIDS should also be interpretable to human experts for reaction and improvement. 
Scalability should also be considered for extending the usage for other cybersecurity tasks such as Android malware detection with little modification. 
Therefore, we summarize our design goals as follows. 

\begin{packeditemize}
    \item \emph{Effectiveness.} The NIDS should be able to achieve a high performance such as accuracy score on detecting the cyberattack category. 
    \item \emph{Privacy-preserving.} The NIDS should be able to protect the data privacy for the network traffic data owners. 
    \item \emph{Interpretability.} The inference process of NIDS can be understood by human experts for further improvements.
    \item \emph{Scalability.} The NIDS should be scalable for different cyberattack detection tasks with little modification on system architectures and model structures. 

\end{packeditemize}

Besides, there are other requirements that are considered as our design goals. 
For instance, the efficiency of the NIDS should be guaranteed for both the ML model training and inference process. 
Then, considering the novel privacy leakage attack on various FL frameworks (e.g. DLG attack~\cite{deep_leakage}), further protection for the data privacy against such gradient recovery attacks must be included.

\section{Design and Implementation}

In this section, we present the design details of \Fname. We give a system overview in Section~\ref{subsec:designoverview} and the description of each step in the rest sections.

\label{sec:approach}
\begin{figure*}[htbp]
    \centering
    \includegraphics[width=0.8\linewidth]{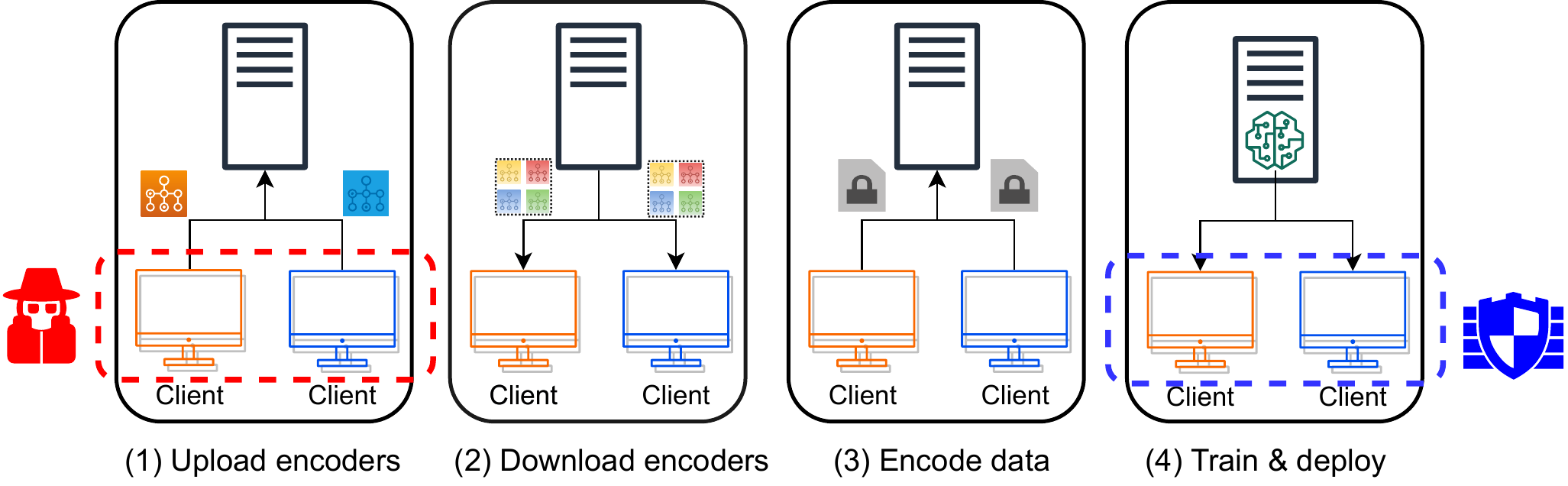}
    \caption{Workflow of the \Fname framework. }
    \label{fig:workflow}
    \vspace{2mm}
\end{figure*}

\subsection{System Overview and Settings}

\label{subsec:designoverview}
\figurename~\ref{fig:workflow} illustrates the workflow to build \Fname. 
There are four steps between the server end (security service provider) and the clients (cyberattack data owners). 
In the first step (\figurename~\ref{fig:workflow}~(1)), the server schedules the involved clients according to the performance requirement, data feature distribution and the cost budget. 
Then, each selected client trains a local encoder (GBDT classifier) with their sampled private data and then uploads encoders containing the local data feature to the server. 
The second step (\figurename~\ref{fig:workflow}~(2)) is to select the most necessary encoders and distribute them to all clients. 
The encoder selection on the server can further reduce the transmission and computation cost. 
Thirdly, each client encodes its own data with the received encoders and sends back the encoding as well as label data to the server (\figurename~\ref{fig:workflow}~(3)). 
The encoding contains necessary features that suffice to identify the attack. As an illustrative example, we can regard data as a cylinder and the encoders as cameras of different positions. The encoding data are pictures taken from different angles.
In the last step, the server trains a global model for final predictions (\figurename~\ref{fig:workflow}~(4)). 
our method does not involve an additional change to the model implementation, which makes our approach easy to deploy and can benefit from existing highly-optimized implementations. 
Overall, we leverage the idea of a partial encoder and global classifier: local features are extracted through encoders and are fed into a global classifier for final prediction.

\subsubsection{Client Setting}
In this paper, we consider only the Horizontal FL setting since the client devices record data of the same feature space, \eg network traffic data recorded by Wireshark. 
Suppose there are $N$ participant clients $P_1,\dots,P_N$. 
The $i$-th client $P_i$ holds data $\mathbf{X}_i$ with label $\mathbf{y}_i$. 
Under Horizontal FL, the client data have the same feature-length, $\mathbf{X}_i\in\mathbb{R}^{n_i\times m}$, where $n_i$ is the number of samples and $m$ is the size of feature space. 

\subsubsection{Data Privacy}
To protect client data privacy against attacks, we adopt random data masking~\cite{super_framework} on the client's training data, \ie randomly drops a certain amount of data, making it more difficult for attackers to recover original data.
Since our data contain two parts, \ie the feature data $\mathbf{X}$ and the label data $\mathbf{y}$, they are of a different kind and need to be treated with different masking methods. 

Formally, the data masking $M$ has the same size as data $\mathbf{X}$ and samples from Bernoulli distribution $Bernoulli(p)$, where $p$ is a hyperparameter to quantify data noise. 
We replace entries of index $(i,j)$ by NaN values to denote lost data if $M[i,j]=1$. 
Remark that data masking is similar to the Dropout\cite{dropout} for the NNs layer. 
As for label data masking, we randomly select $q (0< q< 1)$ labels of $\mathbf{y}$ and replace the selected labels with random different ones.
We summarize the notations in \tablename~\ref{tab:notation}.

\begin{table}[htbp]
\caption{Summary of notations.}
\label{tab:notation}
\centering
\begin{tabular}{cc}
\toprule
\textbf{Notation} & \textbf{Description}              \\ \midrule
$N$               & The number of participant clients \\ \hline
$P_i$             & The $i$-th participant client  \\ \hline
$\mathbf{X}_i$    & The feature data of $P_i$         \\ \hline
$\mathbf{y}_i$    & The label data of $P_i$           \\ \hline
$m$               & The number of data features      \\ \hline
$n_i$             & The number of data samples on $P_i$  \\ \hline
$p$               & Feature data noise paramete \\ \hline
$M$               & Feature data masking matrix  \\\hline
$q$               & Label data noise parameter       \\\hline
$\epsilon$               & Differential privacy budget   \\
\bottomrule
\end{tabular}
\end{table}

\subsection{Step 1: Data Selection \& Upload Encoders}
\subsubsection{Data Selection}
In case there are redundant data among clients, it is essential to select clients in order to improve training efficiency. 
We propose to select clients in a greedy manner based on their own attack type distribution to keep each data type balanced in the training set. 
Concretely, a survey of data type distribution for each participant client is conducted before the main process. 
The survey includes the types of attack and the number of data samples for each type of attack on each participant client. 

Denote $B_{t}$ is the budget of training data samples, which is an important factor of training complexity. 
The objective of this selection step is to select clients whose label data are as much balanced as possible and whose data samples are fewer than $B_t$. 

Formally, let $S_{1},\cdots,S_{k}$ be the $k$ selected clients. 
We denote $N^i_c$ the number of data samples of class $c\in\{1,\cdots,C\}$ belonging to the $i$-th selected client $S_i$. 
Remember $C$ is the number of total attack types. 
Let $N_c = \sum^k_{i=1}N^i_c$ be the number of data samples of class $c$ belonging to the $k$ selected clients and denote $\mathbf{N}=[N_1,\cdots,N_c]$. 
The selected data should be balanced and the amount should be close to $B_t$. 
We use the variance $Var(\mathbf{N})$ in Eq.~\ref{eq:var} to quantify data distribution balance.
\begin{equation}
    \label{eq:var}
    Var(\mathbf{N}) = Var([N_1, \cdots, N_C]) = \frac{\sum^C_{i=1}(N_i -\overline{\mathbf{N}})^2}{C}
\end{equation}



We solve the above problem in a greedy way as shown in Algorithm~\ref{alg:data_selection}. 
In each step, we select the client $P_{i_{min}}$ that can minimize $Var(\mathbf{N})$ and satisfy the training budget $B_t$ (lines 4-10), until there is no clients left (line 17) or no more eligible clients (lines 11-13).

\SetKwInput{KwParam}{Parameters}
\SetKwRepeat{Do}{do}{while}
\SetKw{Break}{break}

\begin{algorithm}[htbp]
    \caption{Data Selection}
    \label{alg:data_selection}
    \SetNoFillComment
    \KwIn{Participant clients $P_1,\cdots,P_N$ and $\{N_c^i\}_{1\leq c\leq C, 1\leq i\leq N}$. Denote $\mathbf{N}^i = [N_1^i,\cdots, N_C^i]$ as data type distribution of participant client $P_i$.}
    \KwOut{Selected participant client $P_{select}=\{S_1, \cdots, S_k\}$.}
    \KwParam{Budget of training data $B_t$.}
    \BlankLine
    \tcc{Initialization}
    $i_0\leftarrow \arg \min_{i\in [\![ 1, N]\!]}Var(\mathbf{N}^i)$,
    $\mathbf{N} \leftarrow \mathbf{N}^{i_0}$,
    $P_{select} \leftarrow \{P_{i_0}\}$, 
    $P_{all} \leftarrow \{P_1,\cdots,P_N\}\backslash\{P_{i_0}\} 
    $;
    \BlankLine
    \Do{$P_{all}\neq \emptyset$}{
        \tcc{Find next participant client in greedy manner.}
        $MinVar \leftarrow Inf$

        \For{$P_i$ in $P_{all}$}{
            $ tmp \leftarrow \mathbf{N}+\mathbf{N}^i$
            
            \If{$Var(tmp) < MinVar \land  Sum(tmp)\leq B_t$}{
                $MinVar\leftarrow Var(tmp)$
                
                $i_{min} \leftarrow i$
            }
        }
        
        \If{$i_{min}$ is None}{
        \Break
        \tcc{No more eligible participant client.}
        }
        
        \tcc{Update parameters.}
        
        $\mathbf{N} \leftarrow \mathbf{N} + \mathbf{N}^{i_{min}}$
        
        $P_{all} \leftarrow P_{all} \backslash
        \{P_{i_{min}}\}$;
        
        $P_{select} \leftarrow P_{select} \cup  \{P_{i_{min}}\}$;
     }

    \KwRet{$P_{select}$}
\end{algorithm}

\subsubsection{Upload Encoders}
For selected clients $S_1, \cdots, S_k$, each client $S_i$ generates an encoder based on its data. 
The trained encoder is composed of a series of decision trees and each inner node of a decision tree represents a decision threshold of a feature. 
Therefore, the encoder does not contain original data nor reversible information. 
Furthermore, the structure of the decision tree (\figurename~\ref{fig:gbdt_example}) makes it possible for human experts to understand the prediction process for further improvement. 
These selected clients then upload trained encoders to the server. 
 
\subsection{Step 2: Encoder Selection \& Encoder Download}

The dimension of training data for the model at the server end (encoding data) is dependant on the number of clients and class partition. 
Therefore, the number of clients that participate in the training has a major influence on the training complexity. 
In order to guarantee the training efficiency, we propose the following heuristic to select the encoders\footnote{For cases with a limited number of clients, all clients will download the encoders to maximize the performance.}.

\subsubsection{Encoder Selection}
The server selects $M$ encoders $(E_i)_{1\leq i \leq M}$ to distribute. 
These encoders are the minimum encoder combination that covers all the classes $\{1,\cdots,C\}$. 
Note that ``the encoder $F$ covers class $c$'' means that $F$ is trained with data containing class $c$.
For example, suppose $C=5$ and the covered classes of each encoder are $\{1,2\}$, $\{2,3,4\}$, and $\{3,4,5\}$. 
The selected encoders in this case are the first one and the third one, as they are the combination of minimum number of encoders to cover all the classes (class $1$ to $5$). 
In summary, the selected encoders should meet the following requirements:

\begin{packeditemize}
    \item $\forall c \in \{1,\cdots,C\}$, $\exists i \in \{1,\cdots,M\}$ such that $F_i$ is trained with data containing class $c$.
    \item $\forall i \in \{1,\cdots,M\}$, the selected encoder combination $\{F_j\}_{1\leq j \leq M} \backslash  \{F_i\}$ can not cover all the $C$ classes. 
\end{packeditemize}

A greedy algorithm is adopted here to select the encoders as shown in Algorithm~\ref{alg:encoder_selection}. 
For each loop, we select the encoder $E_{best}$ that covers the most classes (lines 4-9), and we update the temporary parameters until the selected encoders $\mathbf{E}_s$ cover all the attack classes $\mathbf{C}$ (lines 10-13). 
Note that we can also select encoders trained with potential zero-day attack data in this step. 
For example, during the encoder selection, the server selects not only encoders that cover all the classes but also encoders that contain potential zero-day attack types. 
Such a case is briefly evaluated in Section~\ref{sec:classACC}.

\subsubsection{Encoder Download}
The server sends selected encoders back to the $k$ previously selected clients to control the training complexity. 
The risk of privacy leakage also decreases as the exposure of the potential attack targets, \ie the fewer clients selected, the less possibility for potential information leakage.

\subsection{Step 3: Encode Data}
The clients encode their data by concatenating the softmax probabilities of encoders given their own data. 
Specifically, the softmax probabilities of encoder $F_i$, given $n_S$ samples of the selected participant client $S$ is of size $n_S\times h_i$, where $n_{S}$ is the number of samples on $S$ and $h_i$ is the number of covered classes of the encoder $F_i$. 
Since the sum of softmax probabilities equals one, there is no information lost after deleting one entry of softmax probabilities, \ie it only needs to keep the first $h_i - 1$ columns. 
For example, in the case of binary classification, $C=2$, and it only needs to keep only one of the two columns.

The final size of encoding on the selected participant client $S$ should be $n_{S}\times \sum^M_{i=1} (h_i - 1)$. The participant client then sends the encoding as well as label data back to the server for the final training step. 

To achieve better privacy protection, we apply the Laplace Mechanism~\cite{dwork2006calibrating} on the encoding data. 
Specifically, the sensitivity is $2$, as the encoding data are softmax probabilities. 
Each clients adds noise of Laplace distribution of scale $\frac{2}{\epsilon}$ (see Eq.~\ref{eq:lap_mec}) before sending to the server. 
Other privacy-enhancing techniques, \eg homomorphic encryption, are also applicable in our framework. 
We will discuss this point and list it as future works in Section~\ref{sec:discuss}.

\SetKwInput{KwParam}{Parameters}
\SetKwRepeat{Do}{do}{while}

\begin{algorithm}[]
    \caption{Encoder Selection}
    \label{alg:encoder_selection}
    \SetNoFillComment
    \KwIn{a group of encoders $E_1,\cdots,E_k$, where the class set of $E_i$ is $C_i$.}
    \KwOut{$M$ selected encoders $\mathbf{E}_s$}
    \KwParam{Set of all classes $\mathbf{C} = \{1, \cdots, C\}$.}
    
    \BlankLine
     
    \tcc{Initialization}
    $\mathbf{E}_s = \emptyset, C_s = \emptyset, \mathbf{E}_{left} = \{E_1,\cdots,E_N\};$
    \BlankLine
    \Do{$\mathbf{C} \neq \emptyset$}{
        \tcc{Find next encoder in greedy manner.}
        $MaxOverlap \leftarrow 0$
        
        \For{$E_i$ in $\mathbf{E}_{left}$}{
            \If{$|\mathbf{C} \cap C_i| > MaxOverlap$}{
                $E_{best} \leftarrow E_i$
                
                $MaxOverlap \leftarrow |\mathbf{C}\cap C_i|$
            }
        }
        \tcc{Update parameters.}
        $C_s \leftarrow C_s \cup C_{best}$;
        
        $\mathbf{C} \leftarrow \mathbf{C} \backslash C_{best}$;
        
        $\mathbf{E}_{left} \leftarrow \mathbf{E}_{left} \backslash
        \{E_{best}\}$;
        
        $\mathbf{E}_s \leftarrow \mathbf{E}_s \cup \{E_{best}\}$;
     }

    \KwRet{$\mathbf{E}_s$}
\end{algorithm}

\begin{algorithm}[]
\caption{\Fname Training Phase}
\label{alg:federforest}
\SetNoFillComment
\KwIn{$N$ participant clients $P_1,\cdots,P_N$ with private date $x_1,\cdots,x_N$ and corresponding labels $y_1,\cdots,y_N$}
\KwOut{a joint trained model $F$ composed of a set of encoders and a classifier}
\KwParam{$B_t$ training budget. $\epsilon$ privacy budget.}

\BlankLine
 
\tcc{Step 1.1: Data Selection}
    $P_1,\cdots,P_k \leftarrow DataSelection(P_1,\cdots,P_N, B_t)$

    \For{$i$ in $1,\cdots,k$}{
        $E_i = GBDT(x_i,y_i)$
    }
 \tcc{Step 1.2: \textbf{Send} $E_1,\cdots,E_k$ to server}
 \tcc{Step 2.1: Encoder Selection}
 $\mathbf{E}_s = EncoderSelection(E_1,\cdots,E_k)$
 
 \tcc{Step 2.2: Download encoders to clients}
 \tcc{Step 3: Data Encoding}
 \For{$1\leq i\leq k$}{
        $x_{i_{enc}} = \mathbf{E}_s(x_i) + Lap(\frac{2}{\epsilon})$ 
        \tcc{Encode and send $(x_{i_{enc}},y_i)$ to server}
 }
 $(X_{enc},Y) = \cup_{i=1}^{N} (x_{i_{enc}},y_i)$\\
 $Cls = GBDT(X_{enc},Y)$
 
\KwRet{$F = Cls \circ E_s$}

\end{algorithm}

\subsection{Step 4: Train \& Deploy}
After the server has obtained the encoding and label data, it trains the server classifier (GBDT model) for final predictions. 
When deploying, the service provider needs to combine the selected encoders and the server classifier. The inference consists of two steps: the encoders firstly generate softmax probabilities as the encoding, and the server classifier then outputs final predictions using the encoding.
Thus, another advantage is that the GBDT models allow the security service provider to extract logic rules learned by the model because of the natural interpretability of the tree-based machine learning model.


\section{Experimentation and Evaluation}
\label{sec:experiments}
In this section, we first show the evaluation of experiments on four cybersecurity datasets. Then, We present ablation studies of \Fname to investigate the robustness as well as the performance under different FL scenarios.

\subsection{Experimentation Setup}

The experiments are conducted on a machine installed Ubuntu 20.04 LTS with one CPU AMD Ryzen Threadripper 3960X and 96GB memory. 
We adopt the official FL simulation environment used in competition\footnote{https://1quadrant.com/competitions/competition/detail/5/}.
This simulation environment comprises the client module and the server module, and the training process is the same as FedAvg~\cite{federated_learning}.

\subsubsection{Datasets}
We conduct experiments on the following datasets: competition dataset which is sampled from CIC-DDoS2019\cite{cicddos19} (DDoS2019), CICMalDroid2020\cite{maldroid} (MalDroid2020), CIC-Darknet2020\cite{darknet} (Darknet2020), CIRA-CIC-DoHBrw-2020\cite{DoH} (DoHBrw2020). 
The datasets DDoS2019, Darknet2020, and DoHBrw-2020 contain traffic packet statistic data, \eg flow rate and mean of packet size \etc of network for DDoS attack classification, darknet traffic, and DNS-over-HTTPS (DoH) traffic. 
The Maldroid2020 is an Android malware classification dataset containing static and dynamic analysis patterns for 11,598 Android samples. 

For the dataset DDoS2019, we select 3,355,966 samples as the training dataset and the rest 1,127,526 samples as the test dataset\footnote{This data partition is the same as the competition setting.}. 
Then, in order to prove that \Fname is able to detect the unknown attack as well, we keep the attack ``PORTMAP'' only in the test dataset which means such an attack is not included in the NIDS training process. 
As for the rest three datasets, we randomly take $30\%$ data as test data. The details of the datasets are shown in \tablename~\ref{tab:datasets}. 

\subsubsection{Data preprocessing: masking \& partition}
As mentioned in Section. \ref{sec:problem}, we adopt the following data masking for privacy protection:
\begin{packeditemize}
    \item Random feature data masking with $p=0.1$.
    \item Random label noise with $q=0.2$.
\end{packeditemize}

\begin{figure}[htbp]
    \centering
    \includegraphics[width=0.85\linewidth]{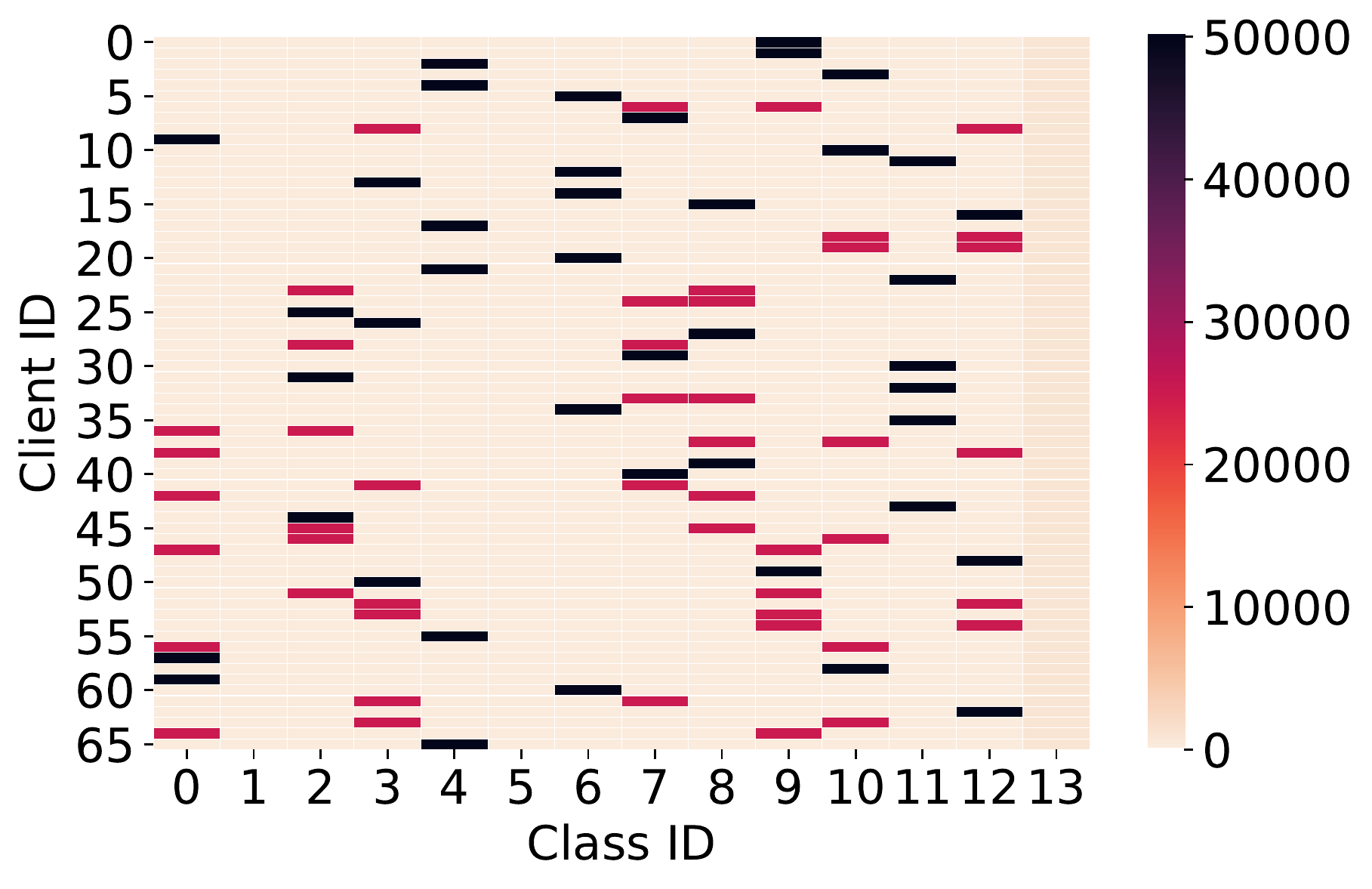}
    \caption{Class distribution of dataset DDoS2019.}
    \label{fig:ddos_distribution}
\end{figure}

The training data are divided into 66 disjoint parts. 
Each part represents data held by a client of the FL setting.
We show class distribution of the competition dataset (DDoS2019) in Fig~\ref{fig:ddos_distribution}. Each client data contain benign data and at most two different types of attack. 

For other datasets, the client data are partitioned in a homogeneous way, \ie data of different classes are equally distributed, with the goal to investigate whether heterogeneous data partition would affect the performance.

\begin{table}[!ht]
\centering
\caption{Statistics of the four datasets.}
\label{tab:datasets}
\begin{tabular}{cccc}
\toprule
\textbf{Dataset}                           & \textbf{\# Features}  & \textbf{\# Class} & \textbf{\# Samples} \\ \hline
DDoS2019                               & 79                    & 14                & 4,483,492  \\ \hline
MalDroid2020                            & 470                   & 5                 & 11,598             \\ \hline
Darknet2020                            & 76                    & 11                & 141,483              \\ \hline
DoHBrw2020                       & 29                    & 4                 & 1,428,407                 \\ 
\bottomrule
\end{tabular}
\end{table}

\subsubsection{Baseline model reproduction}
We choose FedAvg~\cite{federated_learning} with MLP models of $3$, $5$, and $7$ layers as our baseline models. Each hidden layer of MLP has $64$ neurons, and the output layer is equal to the number of classes. We also compare our results with centralized training approaches including XGBoost\cite{xgboost} (noted as X) and 
LightGBM\cite{lightgbm} (noted as L) which can be seen as the best performance achieved by GBDT if we gather client data to the server. 

In our implementation, since MLP models are sensitive to the scale of data as well as infinity (Inf) and missing values (NaN), we first replace invalid values with $0$, then we preprocess the rest data in order to stabilize the training:

\begin{packeditemize}
    \item Logarithm operation. To reduce the order of magnitude of training data $\mathbf{X}$, for $i$-th column $\mathbf{X}[i]$, $\mathbf{X}[i]\leftarrow \log(\mathbf{X}[i]-\min(\mathbf{X}[i])+1)$, where $\min(\mathbf{X}[i])$ is minimum value of $\mathbf{X}[i]$.
    \item Standardization. For $i$-th column $\mathbf{X}[i]$, $\mathbf{X}[i]\leftarrow \frac{\mathbf{X}[i] - mean(\mathbf{X}[i])}{std(\mathbf{X}[i])}$, where $mean$ and $std$ stand for mean and standard deviation value of the column, respectively. Note that this preprocessing operation does not apply to binary feature.
\end{packeditemize} 

To evaluate the performance of MLP, it is necessary to apply the same preprocessing for test data using the min, mean, and standard deviation values that are calculated for preprocessing training data. 
Remark that computing min, mean, and standard deviation of each column is achievable under FL setting: it only needs to compute the sum, the squared sum as well as the number of samples of client data to compute a global value of mean and standard deviation.

We train each MLP model to achieve the highest accuracy score by gird search of hyperparameters: learning rate $lr\in\{0.1, 0.01\}$, maximum round of training for 1,000, 1,500 and 2,000, client batch size of each round 64, 128, and 256. 
We test SGD and Adam\cite{adam} as optimizers.

We use raw data for GBDT models we used for experiments, \ie XGBoost and LightGBM, as they are not sensitive to the order of magnitude and data noise. 
The most influential hyperparameters are been optimized through grid search: $n\_estimators$ ranges from 30 to 200 with step 10, $max\_depth$ from 4 to 10. As for LightGBM, $num\_leaves$ needs to be adjusted according to $max\_depth$, we set $num\_leaves = \lfloor \frac{2}{3}2^{max\_depth} \rfloor$.

\subsubsection{Metrics}
To evaluate the classification performance, we measure and compare the accuracy score, \ie the ratio of correctly classified samples in the test dataset. 
Another important function of NIDS is the attack detection rate, \ie the ability to precisely distinguish malicious and benign data. 
Here, we use the miss rate, \ie the ratio of attack data that are falsely classified as benign over all the benign predictions. 
We also use the F1 score to evaluate the attack detection performance, as it pays equal attention to the precision and recalls in binary classification tasks. 

\subsection{Comparison with baseline models}

We note ``Client GBDT+Server GBDT'' to clarify the model applied on clients and on the serve. 
For example, ``X+L'' signifies that the client encoder model is XGBoost and the server model is LightGBM. 
For this experiment, we set the budget of training data $B_t$ as the length of the dataset, and select all the encoders for encoder distribution of the second step. 
In the ablation study, we will investigate the influence of our two heuristics on the performance in case the amount of clients becomes too large.

\subsubsection{Classification accuracy}\label{sec:classACC}

We report the accuracy scores in \tablename~\ref{tab:results}. 
Bold numbers are the highest accuracy scores obtained on the datasets under the FL setting. 
We can see that \Fname of architecture ``L+L'' achieves systematically higher accuracy against MLP models under FL settings. 
The accuracy scores are comparable with the centralized training version of GBDT. 
The results demonstrate the superiority over NNs for homogeneous as well as heterogeneous data partition, and that our NIDS is scalable for different tasks. 

Remark that LightGBM model and \Fname with LightGBM achieve higher accuracy comparing with XGBoost model. The reason is that LightGBM is based on leaf-wise tree growth when constructing trees, which empowers \Fname with better data noise tolerance ability. 
Another benefit of leaf-wise growth is the reduction of training complexity and model size. 
Consequently, we use ``L+L'' as the default architecture of \Fname for further analysis. As for comparison, we choose 3-layer MLP  (fastest) and 7-layer MLP (most accurate) for NNs-based NIDS with FedAvg.

\subsubsection{Detection performance}

Since the dataset Darknet2020 contains only different types of darknet traffic instead of attack traffic, we choose the rest three datasets to examine the detection performance. We show the miss rates (left) and F1 scores (right) in \tablename~\ref{tab:detection}. The detection performance of \Fname is close to GBDT model (LightGBM) with centralized training (the ideal best performance) and obtains systematically better detection ability than FedAvg with MLP. Note that the miss rates of MLP with FedAvg on DDoS2019 are much higher, because there is little true benign data in the test dataset (only takes 0.28\%), and the predictions of MLP model contain many false-negative errors.

\begin{table}[!h]
\centering
\small
\caption{The miss rates and F1 scores for FedAvg with MLP,  LightGBM with centralized training and \Fname (L+L).}
\label{tab:detection}
\resizebox{0.95\columnwidth}{!}{
\begin{tabular}{cccc}
\toprule
\textbf{NIDS Model} & \textbf{DDoS2019} & \textbf{MalDroid2020} & \textbf{DoHBrw2020} \\ \hline
FedAvg 3-layer      & 93.24 \% / 12.65  & 15.38\% / 72.86       & 12.56\% / 92.27     \\ \hline
FedAvg 7-layer      & 93.31\% / 12.52   & 28.51\% / 68.04       & 5.86\% / 94.17      \\ \hline
LightGBM            & 4.42\% / 94.31    & 6.55\% / 90.84        & 1.00\% / 99.45      \\ \hline
\Fname (L+L)   & 4.40\% / 94.60    & 7.72\% / 88.59        & 0.71\% / 99.54      \\ \bottomrule
\end{tabular}
}
\end{table}

As for the unseen attack ``PORTMAP'' of DDoS2019 in the test data, both FedAvg (7-layer) locally trained GBDT and \Fname (L+L) can detect the attack with $100\%$ success rate, and classify the attack to ``NetBIOS'' with high confidence. 
Since this category of attack (``PORTMAP'' attack) is not labeled and unknown to \Fname, \Fname classifies them all as the similar category ``NetBIOS''. 
Thus, \Fname can effectively detect unknown attacks to generate anomaly alarms.

\begin{table*}[!ht]
\centering
\small
\caption{Accuracy scores of different FL models on four cybersecurity datasets.}
\label{tab:results}
\begin{tabular}{cccccc}
\toprule
\textbf{NIDS Model}                                                             & \textbf{Architecture} & \textbf{DDoS2019} & \textbf{MalDroid2020} & \textbf{Darknet2020} & \textbf{DoHBrw2020} \\ \hline
\multirow{3}{*}{\begin{tabular}[c]{@{}c@{}}NNs - MLP\\ FedAvg~\cite{federated_learning}\end{tabular}} & 3-layer        & 60.29             & 81.26                 & 55.71                & 75.85               \\ \cline{2-6} 
& 5-layer        & 61.37             & 81.36                 & 55.58                & 76.92               \\ \cline{2-6} 
 & 7-layer        & 61.74             & 80.68                 & 56.78                & 76.93               \\ \hline
\multirow{4}{*}{\textbf{\Fname (Ours)}} & X+X & 66.34 & 89.45 & 85.24 & 78.87               \\ \cline{2-6} 
 & L+X & 66.28 & 89.17 & 85.37 & 79.04 \\ \cline{2-6} 
 & X+L & 65.76 & 88.90 & 85.78 & 79.20 \\ \cline{2-6} 
 & L+L & \textbf{67.03} & \textbf{89.63} & \textbf{86.76} & \textbf{79.63} \\ \bottomrule
\end{tabular}
\end{table*}

\subsection{Result Analysis}
\subsubsection{Quality of feature extraction}
\begin{figure*}[htbp]

	    \centering
		\includegraphics[width=\linewidth]{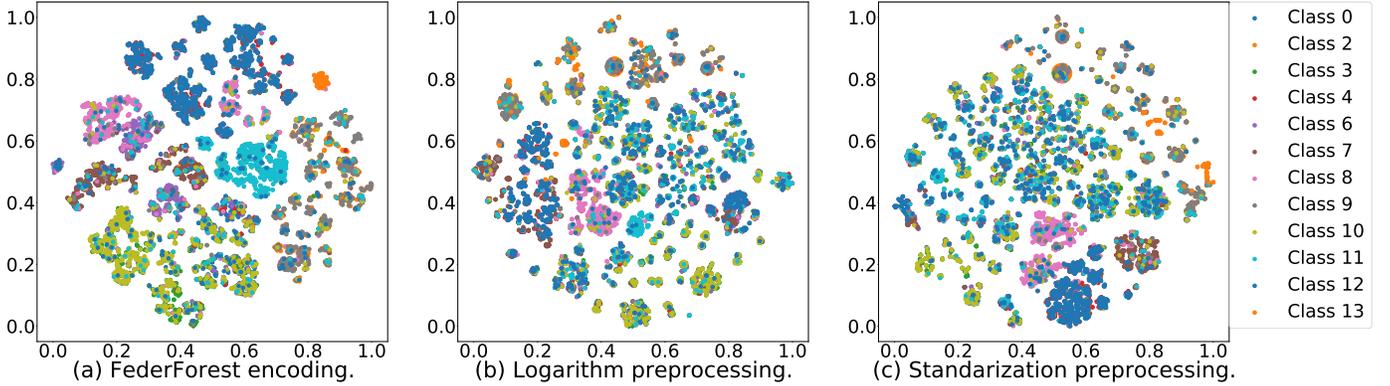}
		\caption{Visualization of encoding data (\textbf{left}) and preprocessed data by log (\textbf{middle}) and standarization (\textbf{right}).}
		\label{fig:visualize}

\end{figure*}

The goal of using encoders in \Fname is to remove noise and private information of raw data and extract feature representations, which can be seen as a method of data preprocessing.
To interpret the performance of our encoders considering the feature extraction, we choose DDoS2019 data as an example and use t-SNE~\cite{tsne} to visualize encoding data (\figurename~\ref{fig:visualize} (a)), as well as data preprocessed by logarithm operation (\figurename~\ref{fig:visualize} (b)) and followed standarization operation (\figurename~\ref{fig:visualize} (c)).

We observe that in the comparison with single logarithm operation, a followed standardization demonstrates better clustering ability for certain classes (\eg~Class 12, blue points), while it still cannot deal with other classes (\eg Class 10, green points) which indicates the weakness of handcrafted preprocessing methods. 
As shown in \figurename~\ref{fig:visualize} (a), the representation of encoding data is less divergent, signifying that encoders outperform others in clustering characteristics for all classes. 
Considering the data and label noise, most samples belonging to the same class are still successfully clustered.
This explains the superiority of \Fname over naive preprocessing techniques for NNs.

\subsubsection{Efficiency Analysis}
Here we use the DDoS2019 dataset as an example to evaluate the training overhead. 
We report in \tablename~\ref{tab:training_overhead} average training overhead required by one client (second row) and the server (third row) for FedAvg using 3-layer MLP and 7-layer MLP and \Fname (architecture L+L).
We observe that the total overhead of \Fname is much lower than NNs when only using CPU for training. 
The reduction is because of the small overhead on the client since the encoders are much smaller than the server classifier. Moreover, most importantly, they do not require updating parameters multiple times like NNs do which proves that \Fname is lightweight and easy to deploy on resource-constrained devices without accelerator like GPUs.

\begin{table}[htbp]
\centering
\footnotesize
\caption{Training overhead of FedAvg (3-layer and 7-layer MLP) and \Fname (L+L) on DDoS2019 dataset.}
\label{tab:training_overhead}
\begin{tabular}{cccc}
\toprule
\textbf{Model} & \multicolumn{1}{l}{\textbf{3-layer MLP}} & \textbf{7-layer MLP} & \textbf{\Fname} \\ \hline 
One Client   & 12.25s                           & 24.83s      & 0.22s             \\ \hline
Server   & 101.12s                          & 241.01s     & 201.26s           \\ \hline
Simulation Time   & 909.30s                           & 1879.78s    & 215.78s           \\ \bottomrule
\end{tabular}
\end{table}

We analyze the communication cost required by NNs-based NIDS and \Fname. 
To quantify communication cost, we measure the transmission data size, \ie the data sent between server and clients during training. 

For the NN-based NIDS model trained by FedAvg, the server sends model parameters to each client and received updated parameters from clients each round. 
Therefore, the communication cost for uploading and downloading are equal, and the complexity is $O(KNR)$, where $R$ is the number of training rounds, and $K$ is the model size and $N$ is the number of participants. 
In our experiment, we set $R=1000$ and $N=66$. 
The model size refers to the weight and bias parameters of MLP, and $K$ equals $14286$ and $30926$ respectively for 3-layer and 7-layer MLPs ($79$ neurons for input layer, $14$ neurons for output layer, and $64$ neurons per hidden layer). 

For \Fname, the uploading and downloading costs are different for the training is different from FedAvg. Specifically, each client uploads trained encoder and encoding data, which takes $O(KN + B_t\times (c_M M +1))$, where $K$ is encoder size, $B_t$ is predefined training data budget, and $c_M$ the average number of classes of encoder among $M$ selected encoders. 
The last $1$ signifies label data. 
The clients download $M$ selected encoders, which demands $O(KNM)$. 
In our experiment, we take all the client data and all the encoders, thus $B_t=3355966$, $M=N=66$ and $c_M M =156$. 
The largest encoder contains $3\times 60$ trees (3 classes and $n\_estimators$ 60) and each tree contains $10$ leaves, thus the encoder size $K$ is smaller than $10 \times 60 \times 3 = 1800$.

We summarize complexity and parameter number for uploading and downloading in \tablename~\ref{tab:communication_overhead}, and transmission data size when each parameter takes $8$ bytes in memory. We observe that FedAvg takes a larger flow quantity than \Fname. More importantly, the downloading data size is much bigger than that of \Fname, which further hinders training speed, since the downloading is generally slower than uploading and the clients need more time to download parameters. 
As for \Fname, the clients only need to download selected encoders for accelerating the total training process. 

\begin{table*}[htbp]
\centering
   \small
\caption{Communication overhead of FedAvg (3-layer and 7-layer MLP) and \Fname (L+L) on DDoS2019 dataset.}
\label{tab:communication_overhead}
\begin{tabular}{cccc}
\toprule
\textbf{NIDS Models}       & \textbf{3-layer MLP}                & \textbf{7-layer MLP}                  & \textbf{\Fname (L+L)} \\ \hline
Uploading Complexity          & \multicolumn{2}{c}{\multirow{2}{*}{$O(KNR)$}}                                & $O(KN  + B_t\times (c_M M+1))$    \\ \cline{1-1} \cline{4-4} 
Downloading Complexity        & \multicolumn{2}{c}{}                                                       & $O(KNM)$                     \\ \hline
Uploading Parameters / Size   & \multirow{2}{*}{942876000 / 7.03GB} & \multirow{2}{*}{2041116000 / 15.21GB} & 526965862 / 3.92GB         \\ \cline{1-1} \cline{4-4} 
Downloading Parameters / Size &                                     &                                       & 7840800 / 7.48MB           \\ \bottomrule
\end{tabular}
\end{table*}

\subsubsection{Privacy Analysis}
Suppose that an honest-but-curious attacker is located on the server-side and aims to recover clients' data. 
The attacker has full control on the server to launch a privacy attack, \eg~DLG~\cite{deep_leakage} uses updated parameters each round to rebuild the client data. 

For \Fname, we claim that it is not achievable to rebuild client data even if the attacker compromises the server (e.g. attacker has the full knowledge of softmax probability and the GBDT model parameter). 
Suppose that there are $C$ classes, and the GBDT model containing $T$ trees for each class, and $n_l$ leaves of each tree. Therefore, the search space for possible logit score combination of classes is $\mathcal{O}(n_l^{CT})$. 
As we demonstrated in experiments, the size of search space is large enough even for the common setting where $n_l=10$, $C=2$, $T=60$. 
Furthermore, as mentioned in Section~\ref{sec:problem}, the client adds random data masking for feature data and label data before training starts, making it more difficult to fully recover the client's data. 
Consequently, except for extremely simple data and models, the attacker can not practically rebuild client data given softmax probability (encoding) and the GBDT model. Moreover, our NIDS can be implemented altogether with privacy-enhancing techniques, \eg homomorphic encryption.

\subsection{Ablation study}
We investigate the impact of parameters on the performance using the competition dataset DDoS2019 as an example.
We use LightGBM as the encoder and classifier for \Fname. 
For comparison, we focus on 3-layer and 7-layer MLP as baseline models. 
Unless otherwise specified, we use the same hyperparameters optimized by grid search as above. 

\subsubsection{Accuracy score (ACC) of different data masking}

As mentioned in Section~\ref{sec:problem}, the clients generate random masking on the feature and label data before the training starts. 
Another benefit is that the clients do not need to transmit the masked data to the server and further reduce communication overhead. 
However, the masked data may contain critical features and decrease the model's performance. 
Particularly, the masking on the feature data is more worth studying since it covers the majority of the data, \ie feature data. 
Thus, we investigate the impact of feature data masking on the model. 
\begin{figure}[htbp]
    \centering
    \includegraphics[width=0.95\linewidth]{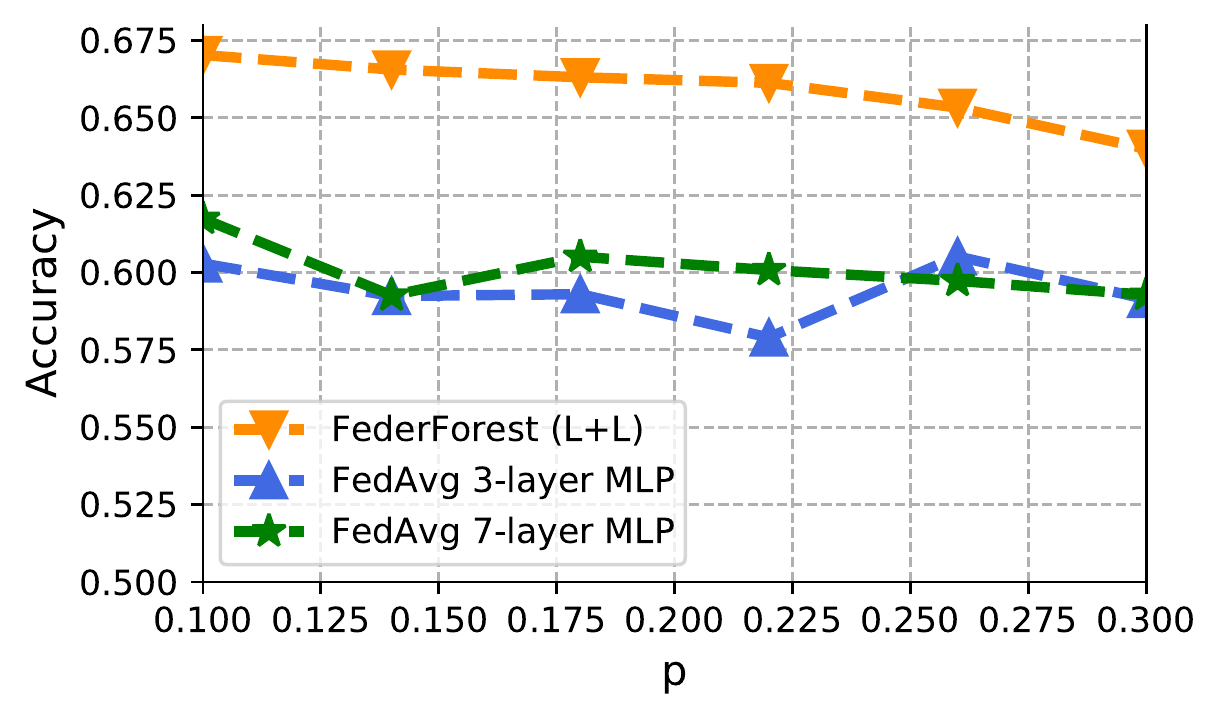}
    \caption{ACC of \Fname and FedAvg for different data masking parameters $p$.}
    \label{fig:data_masking}
\end{figure}

The averaged ACC for different data masking parameters $p$ of \Fname (L+L) and FedAvg (3-layer MLP and 7-layer MLP) are shown in \figurename~\ref{fig:data_masking}. 
Except for $p=0.26$, \Fname achieves systematically higher score. 
We also notice that, comparing with NNs-based model, the performance of \Fname is more sensitive to the change of data masking under the same hyperparameters, as the accuracy score drops larger than 3-layer MLP and 7-layer MLP. 
\subsubsection{Training Budget \& Encoder Selection}
\begin{figure}[htbp]
    \centering
    \includegraphics[width=0.95\linewidth]{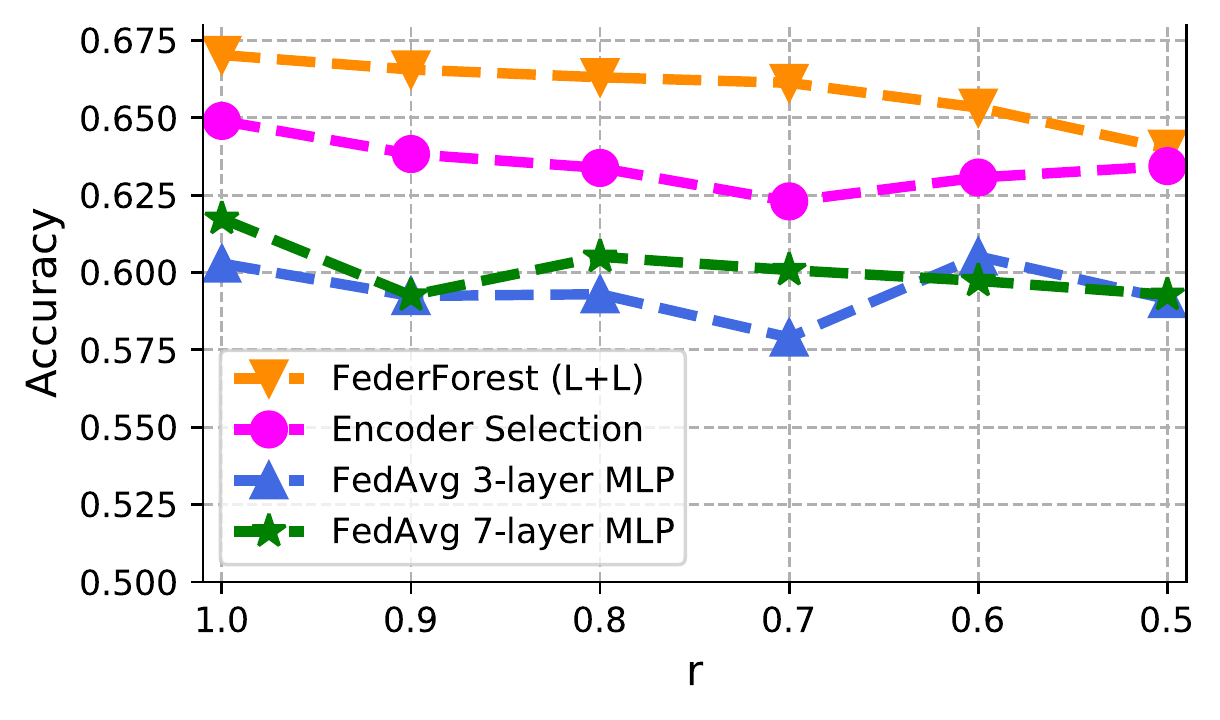}
    \caption{ACC for different training budgets of ratio $r$.}
    \label{fig:data_selection}
\end{figure}

In \figurename~\ref{fig:data_selection}, we present the accuracy for different training budget ratios $r =\frac{B_t}{n}$, where $n$ is the total amount of training data. As illustrated in the figure, the training budget has little influence on the ACC achieved by \Fname, \Fname with encoder selection and FedAvg, as the data are drawn from the same distribution. 

After selected encoders, the size of encoding data is also reduced, therefore it is necessary to readjust the hyperparameters of the global classifier using grid search. Notice that the introduced encoder selection slightly decreases the accuracy score, because there are fewer features in the encoding data used for training the global classifier. However, the encoder selection reduces the communication overhead. In our experiments, at most 8 encoders are selected while there are 66 in total, which means the communication overhead of transferring encoding data is reduced at least 80\%.

\subsubsection{Privacy Budget}

\begin{figure}[htbp]
    \centering
    \includegraphics[width=0.95\linewidth]{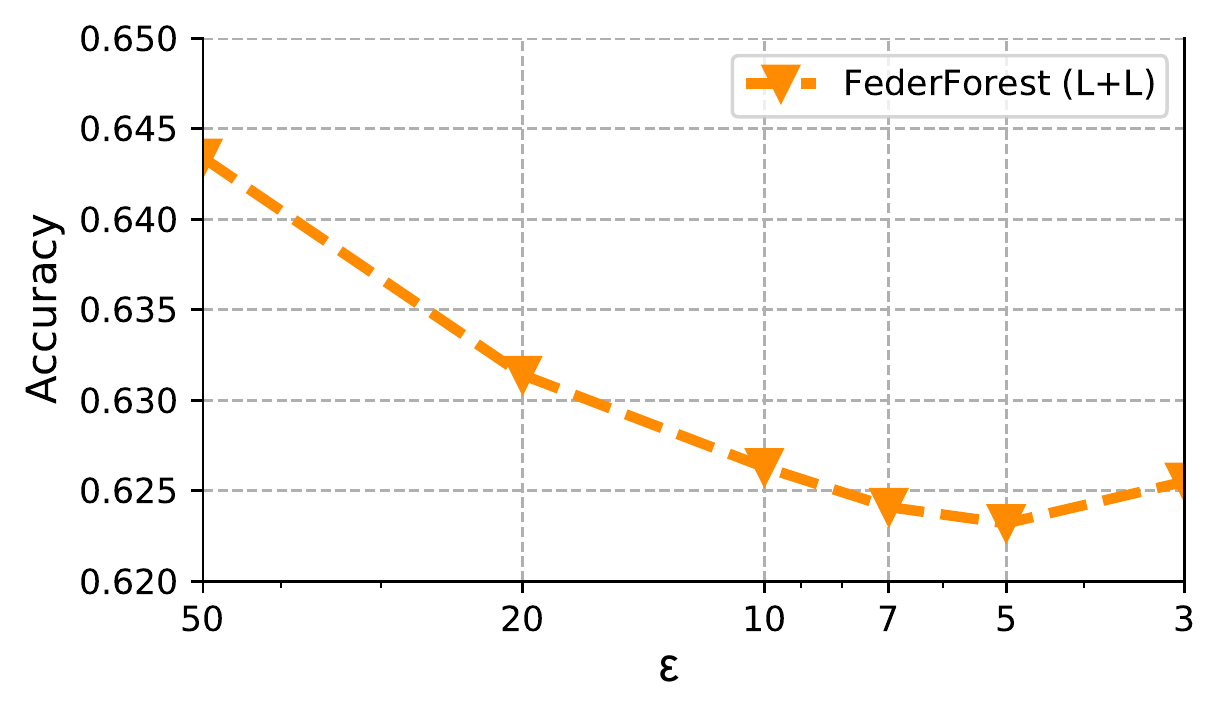}
    \caption{ACC for different differential privacy budget $\epsilon$.}
    \label{fig:dp_epsilon}
\end{figure}

We investigate how differential privacy budget $\epsilon$ affects the accuracy score. In \figurename~\ref{fig:dp_epsilon}, we show the accuracy score of \Fname for $\epsilon\in \{50, 20, 10, 7, 5, 3\}$. Note that smaller $\epsilon$ provides stronger privacy. The accuracy score drops with the budget $\epsilon$, indicating the trade-off between utility and privacy. Still, the scores are higher than NNs-based NIDS without differential privacy protection. We can see that comparing with the other privacy protection measure, \ie random data masking, the differential privacy has a larger impact on the performance of \Fname, as it directly protects the encoding data.

\subsection{Machine Unlearning}

The right to be forgotten attracts increasing attention and is protected by legislations such as the General Data Protection Regulation (GDPR)~\cite{gdpr} in Europe. 
Bourtoule~\etal have introduced machine unlearning \cite{unlearning} which aims to erase the effect of data to be unlearned from a trained model. 
They have also proposed SISA, a training framework that allows data to be unlearned from the model with a minimal cost. 
Specifically, SISA partitions the training dataset into small parts called shards, and train one constituent model for each shard. 
In this way, when some data need to be unlearned, only the model of the shard containing the data has to be retrained from the last checkpoint where the data is not introduced into the training process. 

For the machine unlearning, \Fname is designed to be similar to the SISA by regarding client data as the shards. 
If some client wants to be unlearned, it only needs to delete the corresponding encoding data and retrain the server classifier. 
For the client who wants to be unlearned but whose encoder has been spread to other clients during training, the server has to select another available encoder to encode client data. 
Nevertheless, the major cost is still retraining the server classifier. Consequently, 
\Fname allows participants to freely unlearn data without exerting much influence on the server.





\section{Discussion \& Future work}\label{sec:discuss}

There are three major hyperparameters for training a GBDT model, and six hyperparameters for training \Fname including those for client encoders and those for the server classifier. 
During our experiments, we discover that \Fname is sensitive to the hyperparameters tuning especially for the server classifier. 
Nevertheless, following the official parameter tuning guide of LightGBM and XGBoost, it is possible to determine a search space of common hyperparameters and adopt a grid search to find the optimal. 
In this work, we use grid search over the neighborhood of default hyperparameters. 
Besides grid search, there are various AutoML heuristics that can be applied for searching optimal hyperparameters, \eg random search~\cite{bergstra2012random},  bayesian hyperparameter optimization~\cite{xia2017boosted}. 
We leave these hyperparameter optimization heuristics as future work.

Another issue for building NIDS is to identify the unknown attacks. For those anomaly-based NIDS, although they can alarm when unknown attacks are launched, identification and reacting are still hard since human experts must intervene.
In this paper, we have shown that \Fname has the ability to alarm when unknown attacks exist although the exact category cannot be given. 
In future work, one solution is to train a group of one-vs-rest server classifiers. 
Typically, to train an ensemble of one-vs-rest classifiers, the server needs to train a binary classifier for each attack, and the complexity of total training cost will be linear to the number of attack types. 
In cases where there are many different attack types, a heuristic for reducing the training cost is to manually group the attacks into fewer attack categories. 
Taking multi-class DDoS attack detection as an example, the SYN flood attack and UDP flood attacks belong to the exploitation attack and the other attacks are members of the reflection attack. 
To reduce the attack types, it only needs to train a classifier distinguishing between exploitation attack and reflection attack.



\section{Conclusion}
\label{sec:conclusion}
In this paper, we build \Fname, an NIDS based on a novel collaborative training algorithm by combining FL and GBDT. 
\Fname can effectively detect the cyberattack for different tasks by giving its exact category and zero-day attacks can be easily included in the \Fname framework. 
Moreover, two novel privacy enhancing techniques are proposed to further protect the data privacy in the FL scenario. 
Based on our extensive experimentation, \Fname can provide efficiency, interpretability and scalability for different cyberattack detection tasks which outperforms existing DNNs-based NIDS.

\section*{Acknowledgment}

This work was supported in part by National Key Research and Development Plan of China, 2018YFB1800301 and National Natural Science Foundation of China, 61832013. 

\ifCLASSOPTIONcaptionsoff
  \newpage
\fi

\bibliographystyle{IEEEtran}
\bibliography{ref}




\end{document}